\documentstyle[aps,prl,amssymb,twocolumn,epsfig]{revtex}

\begin{document}

\twocolumn[\hsize\textwidth\columnwidth\hsize\csname@twocolumnfalse\endcsname
\title{Dynamic light scattering from colloidal fractal monolayers}


\author{Pietro Cicuta$^*$ and Ian Hopkinson}

\address{Cavendish Laboratory, University of Cambridge, Madingley Road, 
Cambridge CB3 0HE, U.K.\\ 
$^*$e-mail: pc245@cam.ac.uk}
\maketitle


\maketitle

\begin{abstract}
We address experimentally the problem of  how the structure  of a surface monolayer
determines the visco-elasticity of the interface.
Optical microscopy and surface quasi--elastic light scattering
have been used to characterize aggregation of CaCO$_3$ particles
at the air--water interface. The structures formed by cluster-cluster
 aggregation are two
dimensional fractals which grow to eventually form a percolating
network. This process is measured  through image analysis.  On the same 
system we measure the dynamics of interfacial thermal fluctuations
(surface ripplons), and we discuss how the relaxation process  is affected by the 
growing clusters. We show that the structures start damping 
 the ripplons strongly when the two length scales  are comparable. 
No  
 macroscopic 
surface pressure is measured and this is in contrast to lipid, surfactant or
polymer monolayers at concentrations corresponding to surface
coverage. This observation and the difficulty in fitting the ripplon
spectrum with traditional models suggest that
 a different 
physical mechanism might be responsible for the observed damping of
ripplons in this system.
\end{abstract}




\twocolumn]\narrowtext
\section{introduction}
Surface monolayers of synthetic surfactants, polymers, or
biological molecules like lipids or proteins can dramatically
affect the physical properties of fluid interfaces, in particular
the surface tension and the  elastic and bending moduli
\cite{Lan92}. It is also known that surface properties can be 
modified by the presence of small solid particles \cite{Luc92} At very
 low surface concentrations these macromolecules are, generally,
 in a gas phase. Their
 influence  on the interface properties increases dramatically
 when an overlap concentration
 is achieved. It is of interest to be able to control the
 interface
 parameters, as these in turn determine the  rheology and stability of 
emulsions and foams, which are of
 technological
relevance in different industries from  dairy processing to oil
recovery.\\

Cluster-cluster aggregation is an important class of growth
processes. Particles aggregate into mobile clusters which
 further aggregate between each other to form larger
clusters. In recent years a lot of work has been done to understand the
kinetics of aggregation processes and the resulting structures.
Well studied examples are aggregation of colloids, smoke
particles, carbon black, etc \cite{Mea98}. While many experiments
have been
 performed in 3D, it has been simpler to perform  computer simulations in two
 dimensions. Furthermore, many aggregation phenomena of interest, such as those occurring
 on a surface, are intrinsically two dimensional. For these reasons there has been
an effort to perform experiments also in two dimensions. Some
experiments used the air--water interface to provide a planar
space, and studied first the
    non-diffusive aggregation of  wax balls (diameter
 of the order of a millimeter) \cite{AJ83} and later the diffusion limited aggregation
 of silica microspheres (0.3$\mu$m-diam) \cite{HS85} on the water surface.
\begin{figure}[t!]
           \epsfig{file=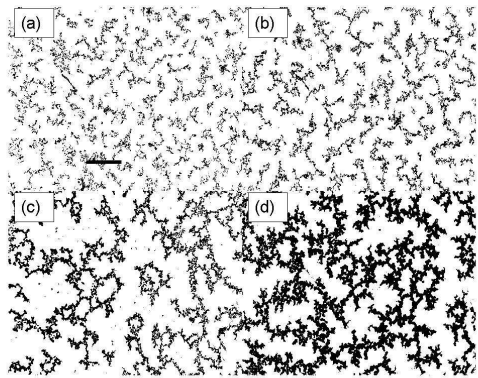,width=8cm}
 \caption{Binarized optical micrographs showing the growth of
fractal aggregates of CaCO$_3$ particles on a water surface. (a)~750s, (b)~990s,
(c)~1830s,(d)~5640s after the creation of a clean surface. The scale bar in (a) is
200$\mu$m; this is comparable to the surface ripplon wavelengths 
that are probed by SQELS.} \label{ima}
\end{figure}
  Further
 experiments created a two dimensional space through confinement between solid
 boundaries. For example polystyrene spheres between glass slides
 (1.1$\mu$m and 4.7$\mu$m -diam) were studied
 under various aggregating conditions\cite{Skj87}. While there  now exists a
theoretical  understanding  of the geometrical structures obtained
from cluster-cluster aggregation, together with detailed
experimental studies, there are few investigations of dynamical and rheological
properties, and, as far as we know, no experiments have been
performed in two dimensions \cite{BH96}.\\

We were motivated to study a system where the  dynamical
interfacial parameters
 could be compared to the geometrical 
structural properties of the surface
 aggregates. A  layer of colloidal particles undergoing aggregation is  an
 ideal choice as it can easily be probed in situ, and such a study would
complement bulk experiments undertaken with a similar motivation \cite{SPS01}.

\section{experiment}
 As a model system we have studied calcium carbonate (CaCO$_3$) particles that form
 at the interface between air and
a solution of calcium hydroxide (Ca(OH)$_2$) as this reacts with dissolved
carbon dioxide. We observe the appearance of micron-sized
particles, which are effectively confined to the interface. These
then aggregate forming two dimensional structures that will be
shown to be fractal.
\begin{figure}[t!]
           \epsfig{file=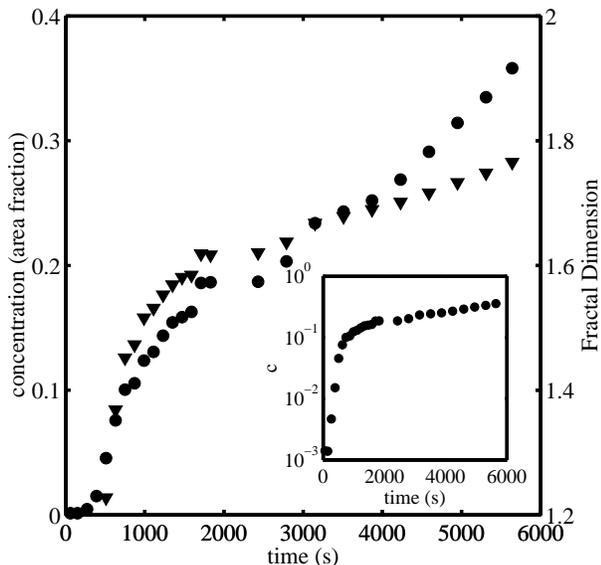,width=8cm}
 \caption{Time evolution of the concentration (the fraction of black pixels in a binarized image)
 ($\bullet$) and the fractal dimension D$_f$ ($\blacktriangledown$) (calculated with the ``box counting'' algorithm). The 
inset shows the same concentration vs. time data on a log-linear scale highlighting the
existence of two separate regimes.} \label{co}
\end{figure}
 This system has been  described  previously \cite{NNM95, WK98}, and in \cite{NNM95} it has
 been shown that the kinetics of aggregation is consistent with fast cluster-cluster
  aggregation  conditions.
 Our experimental conditions are similar to those of
 \cite{NNM95} but we have studied the
system that is obtained from a 0.9g/l calcium hydroxide solution.
At this  concentration, higher than those investigated in  \cite{NNM95} 
(0.1 to 0.3g/l),  the cluster-cluster aggregation
proceeds until a percolating network is formed after about 25
minutes.\\ The samples were prepared by mixing 20ml of 1.2g/l
Ca(OH)$_2$ solution with 10ml of water in a
10cm  diameter petri dish at 23.0$\pm 0.1^\circ$C \footnote{To
avoid the complexities reported in \cite{WK98}, doubly distilled
and deionized water was used to prepare the Ca(OH)$_2$ solution, which
was kept in a glass container and used on the same day. The Ca(OH)$_2$ was
Aldrich 95+\%A.C.S. Reagent.}. Care was
taken to avoid contamination by dust, air drafts and external
vibrations during the experiments. The surface at the center of
the sample was studied with two techniques. First, a time series
of images of the reflected light was recorded with a 1200*1792 pixel digital camera
(Kodak DC290) using
a Zeiss Axioplan microscope, resulting in a resolution of
0.75$\mu$m/pixel and a field of view up to 1.3mm. Second, under
the same conditions, surface dynamic light scattering (SQELS)
measurements  were performed, with an apparatus that  is described
in detail elsewhere  \cite{CH01}. Briefly, SQELS measures the time
correlation of the intensity of light scattered by the ripplons on
the liquid surface, in heterodyne conditions. Surface ripplons are
the thermal fluctuations of a liquid surface, their amplitude is of the order
of a few Angstroms, and their length scale ranges from the molecular to 
the system size. Incident light is
provided by a 30mW HeNe laser, illuminating a region of 5mm
diameter on the liquid surface. The spectrum of the scattered
light has   approximately a Lorentzian form and it can be described in terms
of a frequency and a damping. It is possible in principle to analyze this 
spectrum in detail and  determine the surface
tension, the elasticity of a surface monolayer, and other surface
parameters. In practice this technique, which is described in
detail in \cite{Lan92} and \cite{Ear97},  has been used extensively to study
interfaces of very low tensions and the surface visco--elasticity
of monolayers. With our setup the accessible range of wavevectors
(q) is
  150$<$q$<$500cm$^{-1}$, corresponding to interface roughness
  wavelength
  $\lambda=2\pi$/q between 140 and 500$\mu$m and
frequencies (on water subphase) $15<\omega<100$kHz,  where
$\omega=\sqrt{\gamma q^3/\rho}$, $\gamma$ being  the surface tension
and $\rho$  the density of the liquid subphase. The potential of
this technique for investigating colloid monolayers was first  shown by
Earnshaw \cite{ER90}.\\

\section{results}
In the following we are first going to discuss the characterization
of the system and then the SQELS  measurements of the surface fluctuations.
The digital images have been analyzed consistently
throughout the time evolution using ImageJ and Matlab routines. The first
step was to apply a background subtraction using the ``rolling ball'' algorithm. Then
the images were thresholded at a constant intensity level. These two steps
provided binary images with a white background and the clusters marked as
black.
\begin{figure}[t!]
           \epsfig{file=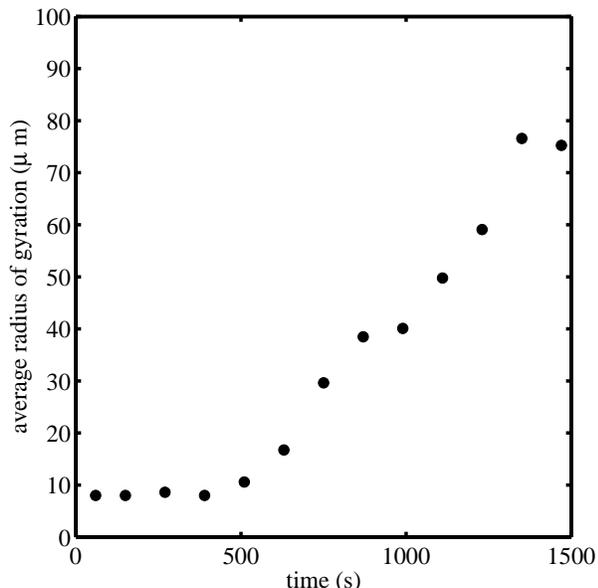,width=8cm}
 \caption{Average radius of gyration. An increase in cluster size is seen 
from 510s. Note that although at  990s the $\langle R_g \rangle$ is only about 40$\mu$m, 
at the same time it can be seen from  Fig.\ref{ima}(b) that the clusters are very elongated and
extend to
 about 200$\mu$m on their principal axis.} \label{newrm}
\end{figure}
  Fig.~\ref{ima} shows  snapshots of the 
aggregating clusters on the
water surface at different times. 
 From these images we  directly calculated the area fraction of
black pixels and the  fractal dimension D$_f$ with the ``box counting'' 
algorithm \cite{Mea98} (box sizes 1,2,3,4,8,12,16,32 and 64 were used). 
The clusters, clearly visible to the human eye, are composed of many closely spaced pixels.
For standard cluster recognition to identify them as a connected structure,
it was necessary to perform a box 7 dilation.  Having labeled pixels as belonging to
separate clusters, it was possible to calculate the average size, mass, number
of clusters for each image.   It is useful to anticipate here that
while the single particles have a  diameter (growing from roughly 2$\mu$m
to 10$\mu$m) which is very small compared to the surface roughness
length scales probed by SQELS,  
the aggregates grow to have a
comparable size and eventually a single cluster spanning more than $1.3$mm
is observed.\\
Fig.~\ref{co} shows the evolution of the concentration (the fraction 
of black pixels) with time. There are two regimes in the concentration increase.
Up to 1000s the increase in concentration is very fast, and new CaCO$_3$
particles are
formed. As they become visible in the  microscope we observe their diameters to be
 between 1 and 3$\mu$m. After 1000s the increase is slower
and we believe it  to be due to growth of the already existing particles. This is consistent
with the estimate, from the images, that the mean 
particle diameter is  around 7$\mu$m at 870s and around 15$\mu$m at 5640s.\\
 Fig.~\ref{ima}(a) shows the surface after
750s,   one can see that aggregation is already occurring and the surface is separated 
in regions of higher and lower particle concentration. Fig.~\ref{ima}(b) shows the surface after
870 seconds when the presence of clusters is very clear.\\
Fig.~\ref{newrm} shows the average radius of gyration $\langle R_g \rangle$ on each image.
  The radius of gyration
 of the i-th cluster is defined as usual as
 $R_{g i}=\sqrt{(1/N)\sum_j{(r_j-r_m)^2}}$, with $N$ the number of pixels, $r_j$ the position of
the j-th pixel and $r_m$ the position of the center of mass. 
The values of $R_g$ on an image are binned into  intervals and the 
 average radius of gyration $\langle R_g \rangle$  is calculated as
$\langle R_g \rangle=\frac{\sum_k{R_{g k}^2\,N(k)}}{\sum_k{R_{g k}\,N(k)}}$, 
where k identifies the bin.  
There is no increase in $\langle R_g \rangle$  up to 510s, and after this it can be seen to
 increase roughly linearly 
with time. This growth cannot be followed beyond
 1500s because of the finite field of view (1.3mm). From 1470s onwards
we observe a structure spanning more than 0.9mm, see Fig.~\ref{ima}(c), while
from 3150s onwards we observe that the  image is  almost entirely composed of a single cluster,
such as in Fig.~\ref{ima}(d).  \\

\begin{figure}[t!]
           \epsfig{file=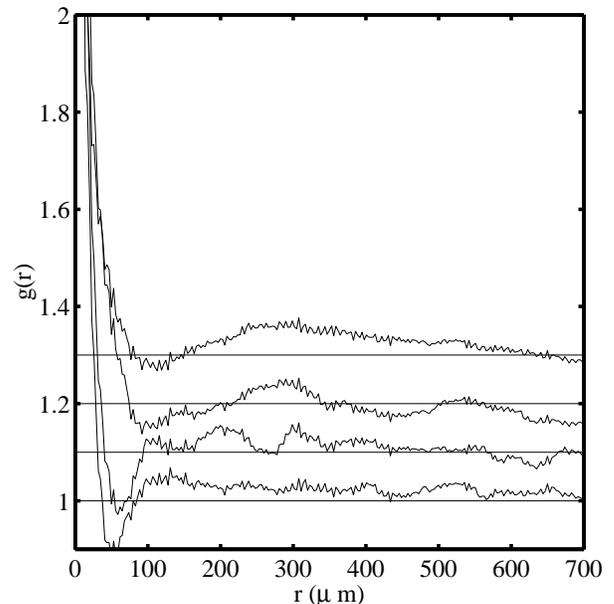,width=8cm}
 \caption{Pair correlation function of black pixels. From bottom to top the times are
 750s, 990s, 1830s, 5640s, corresponding to the images
in Fig.~\ref{ima}. The curves are shifted for comparison and the horizontal
lines are at the expected asymptotic values. The minima in g(r) correspond to the average
cluster radius, these values  are consistent with those from Fig.~\ref{newrm}. The presence of
secondary maxima indicates a loose liquid structure.} \label{gir}
\end{figure}

 Fig.~\ref{gir} shows the pair correlation function g(r) at different
 times.  Clearly, as aggregation proceeds, the position of the minimum 
in g(r) shifts to bigger r  and becomes shallower.  A minimum in g(r) is an anti-correlation, 
indicating that on average across the images the clusters have a 
certain  size. The maximum in g(r) occurs at the average distance between clusters.

\begin{figure}[t!]
           \epsfig{file=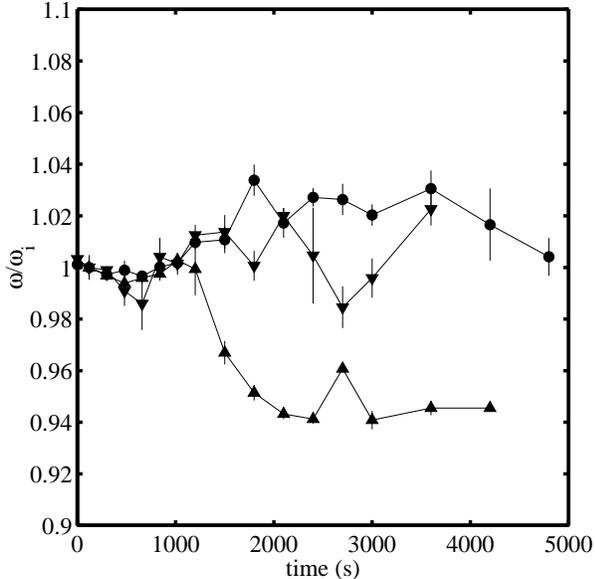,width=8cm}
 \caption{Ripplon frequencies $\omega$,  normalized by their values on  
a clean interface $\omega_i$,  as a function of
time. ($\blacktriangledown$)~corresponds to a  scattering vector
 $q=154.5 \pm 0.2 cm^{-1}$, and  $\omega_i=14.4 \pm 0.1 kHz$;
($\bullet$)~is  $q=222.6 \pm 0.5 cm^{-1}$, and $\omega_i=26.8 \pm 0.2 kHz$;
($\blacktriangle$)~is $q=355 \pm 1 cm^{-1}$, and $\omega_i=53.9\pm 0.4 kHz$. }
\label{omt}
\end{figure}
This  long-ranged correlated structure appears  similar to that observed on colloidal
particles  undergoing   3D cluster aggregation \cite{CG92} and to the surface aggregation
described by Earnshaw in a  2D colloidal system \cite{RE93}.

In Fig.\ref{co} we show the evolution with time of the fractal dimension D$_f$. 
This has been calculated starting from 510s, when the clusters are first present and no deviation
from the expected scaling in the ``box counting'' method was observed. 
 D$_f$ is seen to evolve rapidly from a value of 1.23 to around 1.6 at 1500s and to 
 increase slowly thereafter  as the structures restructure. 
 Restructuring  is observed in the form of free  branches pivoting until
they are doubly connected to the main cluster. This was also observed by \cite{NNM95} and 
is the probable cause for the fractal dimension being higher than that predicted
for simple cluster-cluster aggregation \cite{Mea98}. 

The time correlation functions $G(\tau)$ obtained with SQELS  can be fitted with the form
\begin{eqnarray}
G(\tau)\,=\,B\,+\,A \cos(\omega \tau \,+\,\phi) \exp(-\Gamma \tau)\exp(-\beta^2 \tau^2/4).
 \label{eq1}
\end{eqnarray}
The final Gaussian term is the instrumental broadening, which is calibrated separately 
on a clean liquid surface, and  $\phi$ is a phase term that accounts for the deviation
of the power spectrum from a Lorentzian form \cite{Ear97}. Fitting the data with Eq.\ref{eq1} yields the 
ripplon frequency $\omega$ and damping time
$\Gamma^{-1}$\cite{Lan92}. These parameters  describe the interface
dynamics phenomenologically.
\begin{figure}[t!]
           \epsfig{file=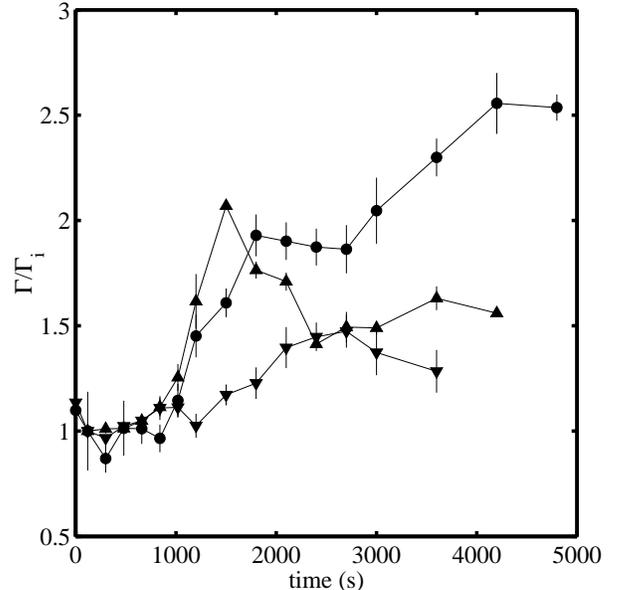,width=8cm}
 \caption{Ripplon damping coefficients $\Gamma$   normalized by their values on  
a clean interface $\Gamma_i$,  as a function of
time. As in Fig.~\ref{omt}, ($\blacktriangledown$)~corresponds to a scattering vector 
$q=154.5 \pm 0.2 cm^{-1}$, and
 $\Gamma_i=0.68 \pm 0.06 kHz$;
($\bullet$)~is $q=222.6 \pm 0.5 cm^{-1}$, and $\Gamma_i=0.73 \pm 0.03 kHz$;
($\blacktriangle$)~is  $q=355 \pm 1 cm^{-1}$, and $\Gamma_i=2.8 \pm 0.1 kHz$. }
\label{gat}
\end{figure}
 No knowledge of the nature of the interface is required
for this analysis, and the fitting procedure is very stable. This is in fact the only meaningful
analysis of the ripplon spectrum that is possible if one does not have a model
that provides a dispersion equation to relate the ripplon spectrum to the  microscopic 
 surface moduli \cite{BJM98}.
We have measured the time evolution of $\omega$ and $\Gamma$  at three different
scattering angles q=155, 223 and 355 cm$^{-1}$, respectively corresponding to  
ripplon wavelengths $\lambda$=405, 282 and 
177$\mu$m and frequencies $\omega$=14.4, 26.8 and 53.9kHz.
For each wavevector we have repeated the experiment three times. In Fig.\ref{omt}
and Fig.\ref{gat} we present the values of  $\omega$ and $\Gamma$ versus time, divided by their
values on a clean water surface, averaged
over the independent  evolutions and binned in appropriate 
time intervals. 
 The values of the frequency, $\omega$, for q=155 and 223cm$^{-1}$ do not vary significantly
from those of the clean water surface. For q=355cm$^{-1}$ there is, after 1200s, a
decrease of about 6\%.  The damping,  $\Gamma$, retains its clean surface value up to 
1000s, when it begins to  increase dramatically for all the wavevectors considered. Up to 1500s it 
can be seen that the increase in damping is bigger for  bigger wavevector. For q=355cm$^{-1}$
  $\Gamma$ has a peak at about 1500s. As noted above, it is not possible from this 
data to determine physical
parameters of the suface without further assumptions, but some conclusions are possible. It is
only after 1000s that the clusters have any effect on the ripplons. This means that it is not
simply the presence of particles but their aggregation into larger structures that 
affects the wave dynamics. At 1000s the diameter of the clusters calculated
 from  the radius of gyration is about 90$\mu$m, but the clusters already  extend above
the capillary wave length scale at least in the direction 
of their principal axis (see Fig.\ref{ima}(b)).\\

With  the same sample preparation, the surface pressure was measured conventionally, 
as a function of time for up to 2 hours,  with a
Wilhelmy filter paper plate. Starting with the clean liquid surface, no pressure increase 
was measured,
even after the formation of the surface layer. This is in contrast to the comparable
situation  occurring with the absorption of surface active molecules on a surface, where the surface 
layer sustains a surface pressure as soon as the molecules 
on the surface form a percolating film. \\

It is important to try to determine the microscopic
physical parameters that give rise to the observed $\omega$ and
 $\Gamma$. This is possible in many cases, as  for example when observing the 
thermal ripplons on a  free liquid surface or in the presence of
 a homogeneous viscoelastic surface monolayer. However the monolayer
in the present study is composed of fractal clusters separated by free liquid surface, 
and it is thus heterogeneous 
on length scales comparable to those of the ripplons. We are not aware of a 
model that takes this into account. It might be that as a first approximation
 to this condition  one should simply  consider the surface as being 
composed of two
kinds of regions (type a and b), with different microscopic parameters,
 each region  scattering light with $\omega_a, \Gamma_a$ 
and   $\omega_b, \Gamma_b$. In this case if $\omega_a$ and  $\omega_b$ are close to each other,
 we would expect to observe an effective broadening
of the scattered power spectrum in time (increase in $\Gamma$), as
 the clusters make the surface
heterogeneous (like Fig.\ref{ima}(b) and (c)), followed by a narrowing (decrease
 in $\Gamma$) as
 the clusters grow to cover the whole
surface evenly (Fig.\ref{ima}(d)). This qualitative behavior of 
$\Gamma$ is indeed observed, in  Fig.\ref{gat} for q=355cm$^{-1}$,  but this cannot be considered
 conclusive of the validity 
of this approximation, since a uniform layer can lead to similar
 behavior of $\Gamma$ (see for example typical data from polymer 
monolayers \cite{Lan92} as a function
of increasing concentration). For lack of a 
 model that takes the heterogeneous quality of the surface into account,  
we have tried  to describe the observed behavior following  the analysis that
is appropriate if a homogeneous
viscoelastic monolayer is present on the surface.
   It is well known \cite{Lan92} that
the dispersion relation
D($\omega$) for waves at an air-liquid interface, bearing a thin
viscoelastic layer, is given by
\begin{eqnarray}
D(\omega)\,=\,\left[\,\epsilon q^2 + i\omega \eta \left(
q + m\right)\, \right] &\times& \nonumber\\ \times \left[\,\gamma q^2 + i\omega\eta
\left( q + m \right) - \frac{\rho \omega^2}{q}\,
\right]  &-& \left[\,i\omega \eta \left(m - q \right)\,
\right]^2,
 \label{eq7}
\end{eqnarray}
where $m\,=\,\sqrt{\,q^2\,+\,i\frac{\omega \rho}{\eta}}$,
$Re(m)>0$,
 $\eta$ is the subphase viscosity, $\rho$ is the subphase
density, $\gamma$ is the surface tension and $\epsilon$ is the
dilational modulus. Solving this equation for D($\omega$)\,=\,0
gives  an expression for the complex wave frequency, $\omega$, as a
function of the scattering vector q. The solutions describe both
dilational and transverse waves. In a light scattering experiment
it is only the transverse waves that scatter light and their power
spectrum P$_q$($\omega$) is given by:
\begin{eqnarray}
P_q(\omega)\,=\,\frac{k_B T}{\pi \omega}Im\left[\,\frac{i \omega
\eta (m\,+\,q)\,+\,\epsilon q^2 }{D(\omega)}\, \right].
 \label{eq9}
\end{eqnarray}
 It is possible to fit 
the correlation function data with the Fourier
transform of Eq.\ref{eq9}, yielding the values of 
$\gamma, \epsilon$ and $\epsilon'$  directly. This
 approach  was first introduced  by Earnshaw (reviewed in \cite{Ear97}) and recently followed
  by ourselves \cite{CH01}.  However in the present work we have not found 
the fit  with three free parameters, nor a fit with the surface pressure fixed to the
independently measured value, to give consistent results.
 This could be because the time evolution of the 
system does not allow for enough data to be acquired,  or because of the inadequacy of 
Eq.\ref{eq7} in 
describing the surface layer. We tried,
in analogy to   many other systems\cite{Mea98} where
  the viscosity  diverges  approaching gelation and an
elastic modulus  develops only  after gelation has occurred,   
 to fit with only $\epsilon '$ as free parameter,
 constraining the surface pressure to the pure
liquid phase value (72.2mN/m) and the real part of the dilational modulus to 0. This  gave
us a dilational viscosity $\epsilon '$ that was initially zero and  increased rapidly
after about 1100s. Until the use of 
 Eq.\ref{eq7} is justified, this approach certainly cannot be considered quantitatively correct.

\section{conclusions}
We were motivated to study a two dimensional  system where the rheological properties could be
related to the structure. We  have been   successful in  characterizing
the aggregation process of CaCO$_3$ particles on a surface
and   measuring the effect of growing clusters on surface  ripplons.
 The data presented in Fig.\ref{omt} and Fig.\ref{gat} describe the 
dynamics of the surface ripplons and show that the clusters begin to modify the wave 
relaxation process
 when their size is of the order of the surface wavelength.
 On the basis of this data it is also clear that in the initial 
stages of the growth, a given cluster size has a damping effect which is stronger
for bigger wavevectors, suggesting that  we are  probing  a
scale dependent gelation, where percolation on the length scale of
a surface wave  has an effect on that ripplon's  dynamics,
 while the macroscopic behavior of the
system is still liquid-like.  These  are new observations.  
It remains to be explained how to relate
the dynamical behavior to the surface layer  moduli. We have suggested what the effect of
a heterogeneous surface could be and we have described the difficulties in fitting
the data with the conventional surface layer model.  We feel that to understand
the behavior shown in  Fig.\ref{omt} and Fig.\ref{gat} a  model is needed to
 specifically take 
into account the fractal
nature of the aggregates and the heterogeneity of the surface as a whole.

\begin{acknowledgments}
We acknowledge useful discussions with Rafi Blumenfeld, Martin Buzza and Eugene Terentjev.
\end{acknowledgments}


\begin{references}

\bibitem{Lan92}
    D.Langevin,
    {\it Light Scattering by Liquid Surfaces and Complementary
    Techniques},
  Dekker,
  New York,
  1992.

\bibitem{Luc92}
   J. Lucassen,
   Colloids and Surfaces"
   {\bf 65},
   139 (1992).


\bibitem{Mea98}
   P. Meakin,
   {\it Fractals, scaling and growth far from equilibrium},
   Cambridge University Press,
   Cambridge (U.K.),
   1998.

\bibitem{AJ83}
   C. Allain and B. Jouhier,
   J. Phys. (Paris) Lett.,
   {\bf 44}
   L421 (1983).

\bibitem{HS85}
   Alan J. Hurd and Dale W. Schaefer,
   Phys. Rev. Lett.
   {\bf 54}
   1043 (1985).

\bibitem{Skj87}
   A.T. Skjeltorp,
   Phys. Rev. Lett.
   {\bf 58}
   1444 (1987).


\bibitem{BH96}
   A. Bunde and S.Havlin (Eds.),
   {\it Fractals and disordered systems (2nd edition)},
   Springer-Verlag,
   Berlin,
   1996.

\bibitem{SPS01}
   P.N.Segre, V.Prasad, A.B.Schofield and D.A.Weitz,
   Phys. Rev. Lett.
   {\bf 86},
  6042 (2001).





\bibitem{NNM95}
   T. Nakayama, A. Nakahara and M. Matsushita,
   J. Phys. Soc. Japan
   {\bf 64},
   1114 (1995).


\bibitem{WK98}
   H.H. Wickman and J.N. Korley,
   Nature
   {\bf 393}
   445 (1998).

\bibitem{CH01}
   P. Cicuta and I. Hopkinson,
   J. Chem. Phys.,
   {\bf 114}
   8659 (2001).

\bibitem{Ear97}
   J.C. Earnshaw,
   Appl. Optics
   {\bf 36},
   7583 (1997).







\bibitem{ER90}
   J.C. Earnshaw and D.J. Robinson,
   J. Phys. Condens. Matter
   {\bf 2},
   9199 (1990).






\bibitem{CG92}
   M. Carpineti and M. Giglio,
   Phys. Rev. Lett.
   {\bf 68},
  3327 (1992).




\bibitem{RE93}
   D.J. Robinson and J.C. Earnshaw,
   Phys. Rev. Lett.
   {\bf 71},
   715 (1993).










\bibitem{BJM98}
   D.M.A.Buzza, J.L.Jones, T.C.B.McLeish and R.W.Richards,
   J. Chem. Phys.
   {\bf 109}
   5008 (1998).








\end{references}

\end{document}